\documentclass[aps,pre,reprint,groupedaddress]{revtex4-1}

\usepackage{graphicx}
\usepackage{color}

\begin{document}


\title{Motility of \textit{Escherichia coli} in a quasi-two-dimensional porous medium}



\author{Juan Eduardo Sosa-Hern\'andez}
\email[]{jsosa@cinvestav.mx}
\affiliation{Cinvestav Unidad Monterrey, Parque de Investigaci\'on e Innovaci\'on Tecnol\'ogica (PIIT), Apodaca, Nuevo Le\'on 66600, M\'exico}

\author{Mois\'es Santill\'an Zer\'on}
\email[]{msantillan@cinvestav.mx}
\affiliation{Cinvestav Unidad Monterrey, Parque de Investigaci\'on e Innovaci\'on Tecnol\'ogica (PIIT), Apodaca, Nuevo Le\'on 66600, M\'exico}

\author{Jes\'us Santana-Solano}
\email[]{jsantana@cinvestav.mx}
\affiliation{Cinvestav Unidad Monterrey, Parque de Investigaci\'on e Innovaci\'on Tecnol\'ogica (PIIT), Apodaca, Nuevo Le\'on 66600, M\'exico}


\date{\today}

\begin{abstract}
Bacterial migration through confined spaces is critical for several phenomena like: biofilm formation, bacterial transport in soils, and bacterial therapy against cancer \cite{Whitman1998, Toley2012, Berdakin2013, Barton1995, Binz2010, Jimenez-Sanchez2012}. In the present work, \textit{E. coli} (strain K12-MG1655 WT) motility was characterized by recording and analyzing individual bacterium trajectories in a simulated quasi-2-dimensional porous medium. The porous medium was simulated by enclosing, between slide and cover slip, a bacterial-culture sample mixed with uniform 2.98 $\mu m$ spherical latex particles. The porosity of the medium was controlled by changing the latex particle concentration. By statistically analyzing trajectory parameters like: instantaneous velocity and turn angle, as well as mean squared displacement, we were able to quantify the effects that different latex particle concentrations have upon bacterial motility. To better understand our results, bacterial trajectories were simulated by means of a phenomenological random-walk model (developed ad hoc), and the simulated results were compared with the experimental ones. 
\end{abstract}


\maketitle

\section{Introduction}

Motility is an organism capability to move in an independent and spontaneous fashion. In general, motility enhances an organism opportunities to succeed on reproduction and growth, and to escape from hazardous environments. Many diseases, biofilm formation, and bioremediation are possible thanks to microorganism motility.

There exist two different types of flagellum-driven motile microorganism: pushers and pullers, depending on whether the driving force is respectively generated ahead or behind the organism main body. In this work we study \textit{E. coli} motility, which is a pusher. This bacterium 3-dimensional motility is well understood when it is freely swimming, and was mostly characterized by the seminal work of H. C. Berg \cite{Berg1972,Berg2004}. \textit{E. coli} swimming consists of alternated runs and tumbles. 
During runs, swimming is persistent and mostly unidirectional. This motility mode is caused by a bundle of flagella rotating clockwise (forward view) and in phase---because of angular momentum conservation, the bacterium body rotates counter-clockwise---and its average duration is about 1 s. Conversely, bacteria constantly re-orientate to a random direction during tumbles. Tumbling occurs when at least one flagellum rotates clockeise, making the bundle break and the flagella spread. The average duration of tumbles is approximately 0.1 s.  

Although the free-swimming 3-dimensional motility of this microorganism has been extensively studied \citep{Berg1972, Berg1995, Wu2006, Turner2000, Turner2010}, as well as its motility near solid surfaces \cite{Berg2004, Lauga2006, Drescher2010, Drescher2011, Giacche2010}, several questions regarding motility under other conditions remain open. For instance, the effects that confined spaces have on motility have recently become a focus area of research, not only for \textit{E. coli}, but also for other microscopic swimmers. This interest emerges form the fact that most microorganisms on Earth live in porous environments, like soils and biofilms \cite{Binz2010, Jimenez-Sanchez2012, Kusy2007, Olson2005, Olson2006}. Examples include studies on synthetic micro-swimmers such as Janus particles \cite{Brown2014, Ghosh2014, Spagnolie2015, Takagi2014, Volpe2011, Zheng2013}, mathematical modeling studies \cite{Brotto2013, Hernandez-Ortiz2005, Jabbarzadeh2014, Lushi2014, Najafi2013, Sandoval2014}, and motility studies within constricted spaces \cite{Lambert2010, Liu1995, Rutllant2005, Wan2008, Kusy2007, Ping2015}. However, bacterial motility in micron and sub-micron constricted spaces has only been studied with ideal geometries \cite{Mannik2009, Berdakin2013, Ping2015, Raatz2015, Jabbarzadeh2014}. In order to study bacterial motility in a more realistic environment, we have generated a device that simulates a quasi-two-dimensional porous media \cite{Davit2013}, and studied how \textit{E. coli} motility is affected by varying porosity conditions. Our aim is to retrieve information that could help us find answers for problems in biology, medicine, and ecology, that involve microbial motility \cite{Mannik2009, Toley2012, Narayanaswamy2009, Kusy2007}.

\section{Materials and methods}

\subsection{Cell culture}

\textit{Escherichia  coli} cell cultures (wild-type strain MG1655) were prepared from frozen stocks according to the experimental protocol reported in \cite{Lauga2006}. First, the bacterial culture was saturated by seeding bacteria in lysogeny broth medium (LB, 10 g/L tryptone, 5 g/L yeast extract, and 10 g/L NaCl), and letting them grow for 16 h at 34 $^\circ$C, shaking at 150 rpm. Next, samples containing 200 ${\mu}$L of the saturated culture, with glycerol at 15 $\%$ ${(v/v)}$, were stored at -80 $^\circ$C. 98 cell culture tubes were prepared from each stock cycle to guaranty consistency between experiments. For each experimental session, a new cell culture was grown from a stock tube by adding 4 mL of fresh LB in 8 mL tubes, containing 200 ${\mu}$L of saturated LB. The resulting cultures were then grown at 34 $^\circ$C for 3.5 h on a rotary shaker (150 rpm), in order to reach the mid-exponential phase of bacterial growth. These conditions lead to an OD600 = 0.98. Experimental samples were prepared by washing 1 mL of bacterial culture and resuspending it in fresh motility buffer (10 mM potassium phosphate, 0.1 mM EDTA, 10 mM NaCl; pH = 7.5). Each sample was washed three times by centrifuging the cell culture at 2000 g for 5 min and dispersing it with fresh motility buffer. Finally, the sample was stored for 15 min, so that dead and non-motile bacteria sediment, and medium from the tube top was used to ensure approximately 90 $\%$ of motile bacteria.

\subsection{Experimental setup}

To confine a dilute suspension of bacteria in a quasi-two-dimensional porous medium, we placed a mixture of bacterial culture and 2.98 ${\pm}$ 0.14 ${\mu}$m-diameter polystirene spheres (Thermo Scientific) on a slide, and gently covered it with a cover slip, following the procedure detailed in \cite{Santana-Solano2005}. The spherical beads were employed to fulfill two different purposes: 1) to act as pillars between slide and cover glass, and 2) to generate a disordered porous matrix in which the beads act as obstacles for bacterial swimming. The obstacle area fraction was controlled by changing the concentration of beads. In our experiments we considered area fraction values ranging from 0.01 to 0.4. In all cases, the bacterial count was kept between 15 and 30 cells within the video field. Evaporation and externally caused fluxes were prevented by sealing the space between glass edges with semi-polymerized polydimethylsiloxane (PDMS). The volume of each sample, prepared as we have just described, is about 8 ${\mu}$L, ensuring a mono-layer of beads. To prevent bacteria from attaching to glass surfaces, slides and coverslips were previously treated with PVP-40 (polyvinylpyrrolidone) in Mili-Q water at 0.005 $\%$ ${(w/v)}$ \cite{Berke2008}. All experiments were performed at 25 $^\circ$C. Videos were recorded (at 30 fps, with a resolution of 480 $\times$ 720 pixels) by means of a CCD camera mounted on an Olympus BX51 microscope, with a 40$\times$ magnification phase contrast objective.  With this setup, the frame dimensions are $120 \times 160$ ${\mu}$m$^{2}$. In all cases, the recorded videos were 5 min long.

\subsection{Trajectory analysis}

Bacterial trajectories were recovered from the recorded videos via an image-analysis algorithm that was originally developed by Crocker and Grier \cite{Crocker1996}, and is implemented on \texttt{MatLab}. Basically, this algorithm identifies individual bacteria in every video frame, and them optimizes an adequate objective function to link bacterium positions in consecutive frames. To characterize the obtained trajectories, we computed the instantaneous velocity ($\vec{v}(t)$) and turn angle  ($|\theta(t)|$), as well as the Mean Squared Displacement (MSD) ($\langle {\Delta \vec{r}(t)}^2 \rangle$), by means of the following equations:
\begin{eqnarray}
\vec{r}(t)=x(t)\hat{x}+y(t)\hat{y}+0\hat{z}, \label{eq01} \\
\vec{v}(t) = \frac{\vec{r}(t)-\vec{r}(t-\Delta{t})}{\Delta{t}}, \label{eq02} \\
\theta(t)= cos^{-1}\left(\frac{{\vec{v}(t-\Delta{t})\cdot\vec{v}(t)}}{|\vec{v}(t-\Delta{t})||\vec{v}(t)|}\right), \label{eq03}\\
\langle {\Delta \vec{r}(t)}^2 \rangle = \langle (\vec{r}(t+n\Delta t) - \vec{r}(t) )^2\rangle, \label{eq04}
\end{eqnarray}
where $\vec{r}(t)=x(t)\hat{x}+y(t)\hat{y}+0\hat{z}$, $\Delta{t}$ is 1/30 s, $n$ is an integer ranging from one to the trajectory length, and $\langle...\rangle$ denotes ensemble average. 

Following Masson \cite{Masson2012}, we measured run and tumble lengths. In our case, runs and tumbles were identified by means of a Schmitt trigger, in which bacterial speed is the threshold parameter. In summary, when a bacterium speed is above 0.65 times the trajectory mean speed, the bacterium is considered to be in a run. On the other hand, if the bacterium speed is below 0.6 times the trajectory average speed, the bacterium is regarded to be in a tumble. By statistically analyzing all the measured run and tumble times, we found that the corresponding probability density functions (PDFs) are well fitted by exponential distributions of the form:
\begin{equation}
    \rho(t) =\frac{1}{\tau} e^{-t/\tau },
    \label{Eq07}
\end{equation}
in which $\tau$ is the average residence time. 

Finally, for each of the bacterial swimming stages: run and tumble, we performed an analysis of velocity components, using the unitary velocity of the previous step as a reference. We obtained the longitudinal ($\vec{v}_{\parallel}$) and transverse ($\vec{v}_{\perp}$) velocity components  as follows:
\begin{eqnarray}
    \vec{v}_{\parallel}  =  \vec{v}(t) \cdot \frac{\vec{v}(t-\Delta t)}{|\vec{v}(t-\Delta t)|},
    \label{Eq05}\\
    \vec{v}_{\perp} = \vec{v}(t) \cdot \left( \hat{k} \times  \frac{\vec{v}(t-\Delta t)}{|\vec{v}(t-\Delta t)|} \right).
    \label{Eq06}
\end{eqnarray}
After carrying out the corresponding statistical analysis, we found that the experimental PDFs for both the longitudinal and the transverse velocity components, for both motility modes (runs and tumbles), are well fitted by normal distributions:
\begin{equation}
    P(x) = \frac{1}{{\sigma \sqrt {2\pi } }}e^{{{ - \left( {x - \mu } \right)^2 } \mathord{\left/ {\vphantom {{ - \left( {x - \mu } \right)^2 } {2\sigma ^2 }}} \right. \kern-\nulldelimiterspace} {2\sigma ^2 }}},
    \label{Eq08}
\end{equation} 
where $\mu$ and $\sigma$ are the distribution mean value and standard deviation, respectively.

\subsection{Mathematical model}

We developed a simple mathematical model to mimic bacterial swimming as a phenomenological random walk. In this model, bacteria were regarded as hard disks (without flagella) with a diameter equal to the average minor semi-axis of an \textit{E. coli} bacterium (1 $\mu$m). The obstacles were represented by 2.98 $\mu$m-diameter hard disks and, for the sake of consistency, they were placed at the same spots the latex beads occupy in the experimental videos. All the performed simulations accounted for ten bacteria swimming in a 160 $\times$ 120 $\mu$m surface, with periodic boundary conditions, and they consisted of 1 million, $1/30$ s long, steps. The pseudo code for the mathematical model is as follows:
\begin{enumerate}
  \item Set the time step $\Delta t = 1/30 \, \text{s}$.

  \item Set the initial simulation time to $t = 0$, and randomly chose the $x$ and $y$ coordinates of the initial trajectory point from uniform distributions in the ranges $[0, 160]$ $\mu$m and $[0, 120]$ $\mu$m, respectively. Take care that the initial trajectory point does not lie within an obstacle.

  \item Randomly select the initial motility mode: persistent or tumbling, considering a probability of 0.5 for each one.

  \item Randomly compute, from the PDF in Eq. (\ref{Eq07}), the time $T$ the simulated bacterium will remain in the current motility mode. To do this, consider the parameter values of the corresponding motility mode (see the Results section).

  \item Calculate the number of steps to be given in the current mode as the integer part of $N = T / \Delta t$.

  \item For every step, randomly calculate the velocity longitudinal and transverse component by means of the PDF in Eq. (\ref{Eq08}), with the corresponding parameter values (see the Results section). Then, multiply times $\Delta t$ to get the corresponding displacement.

  \item For every step, check the resulting virtual position to detect obstacles. If the bacterium collides with an obstacle, randomly chose the collision type (either ``arch'' or ``tangent'', as described below), and use it to calculate the corrected position.

  \item Update the simulation time, $t := t + N \Delta t$.

  \item Change the bacterium motility mode.

  \item Iterate from step 4.
\end{enumerate}

By carefully observing the experimental videos, we were able to identify two types of collisions (here termed ``arch'' and ``tangent'') between bacteria and obstacles. After a bacterium encounters an obstacle, it swims along the obstacle edge if the collision is of the ``arch'' type. Conversely, after a ``tangent'' collision, the bacterium follows a linear trajectory, tangential to the obstacle edge at the contact point. In our model, we simulate these collisions as follows:
\begin{enumerate}
  \item If as the result of a trajectory step, the simulated bacterium would have to penetrate or swim across an obstacle, we allow the bacterium to reach the obstacle edge, and then we make it swim for a given distance $d \leq l$ (with $l$ the length of the original step minus the distance traveled up to the obstacle edge) either along the obstacle edge (for an ``arch'' collision), or tangentially to it (for a ``tangent'' collision). 
  
  \item Each time a bacterium collides with an obstacle, the bacterium rests in touch with the bead surface for a while, before it continues swimming. Based on former studies of bacterium interactions with walls and beads \cite{Li2008, Liu2009, Liu2011, Sipos2015}, we take this into account by computing distance $d$ as $d = \alpha l$, with $\alpha$ a random number uniformly selected in the interval $[0.8, 1.0]$. 

  \item By trial and error we found that assuming the corresponding probabilities of ``arch'' and ``tangent'' as $7/9$ and $2/9$, makes the simulation results concord with the experimental ones. This agrees with reference \cite{Spagnolie2015}, in which the authors assert that the most important contribution to bacteria-obstacle interactions is due to swimming parallel to surfaces.
\end{enumerate}

\section{Results and discussion}

\subsection{Trajectory analysis at very low obstacle concentration}

We started by recording bacterial motility videos (following the procedure detailed in the Materials and Methods section) in which the obstacle concentration was low enough (obstacle area fraction about 0.01) to keep bacterium-obstacle interactions to a minimum, while ensuring a 2.98 $\mu$m separations between glass surfaces. Thereafter, we gradually increased obstacle concentration to evaluate the effect that bacterium-obstacle interactions have upon bacterial motility. The largest obstacle area fraction we achieved was about 0.4. In total, we obtained 88 videos, and at least two of them correspond to each one of the considered area fractions. 

\begin{figure}[htb]
\includegraphics[width=.48\textwidth]{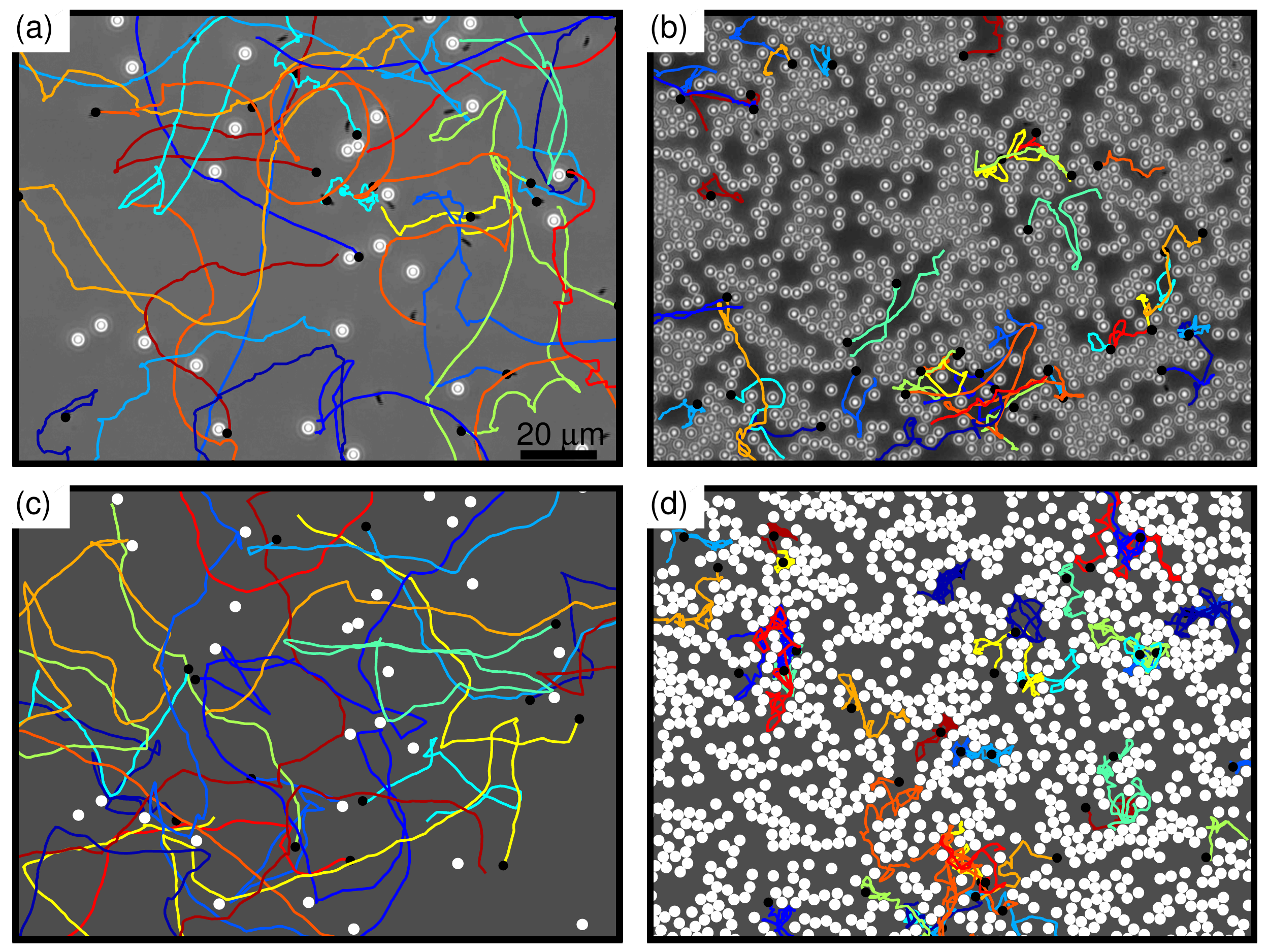}
\caption{Experimental---(a) and (b)---and simulated---(c) and (d)--bacterial trajectories under low---(a) and (c)---and high---(b) and (d)---obstacle-area-fraction conditions. In all cases, the obstacles are represented as white circles.}
\label{Fig01}
\end{figure}

As described in the Materials and Methods section, we recovered all the possible bacterial trajectories from the recorded videos. On average, we were able to obtain 2690 trajectories from every video; the average trajectory length being 46 steps (about 1.5 seconds). In Figs. \ref{Fig01}(a) and (b) we show representative samples of the trajectories we got from typical videos recorded at low (a) and high (b) obstacle area fractions. Notice how both the experimental and the simulated trajectories become more tortuous as the obstacle area fraction increases.

\begin{figure}[htb]
\includegraphics[width=.48\textwidth]{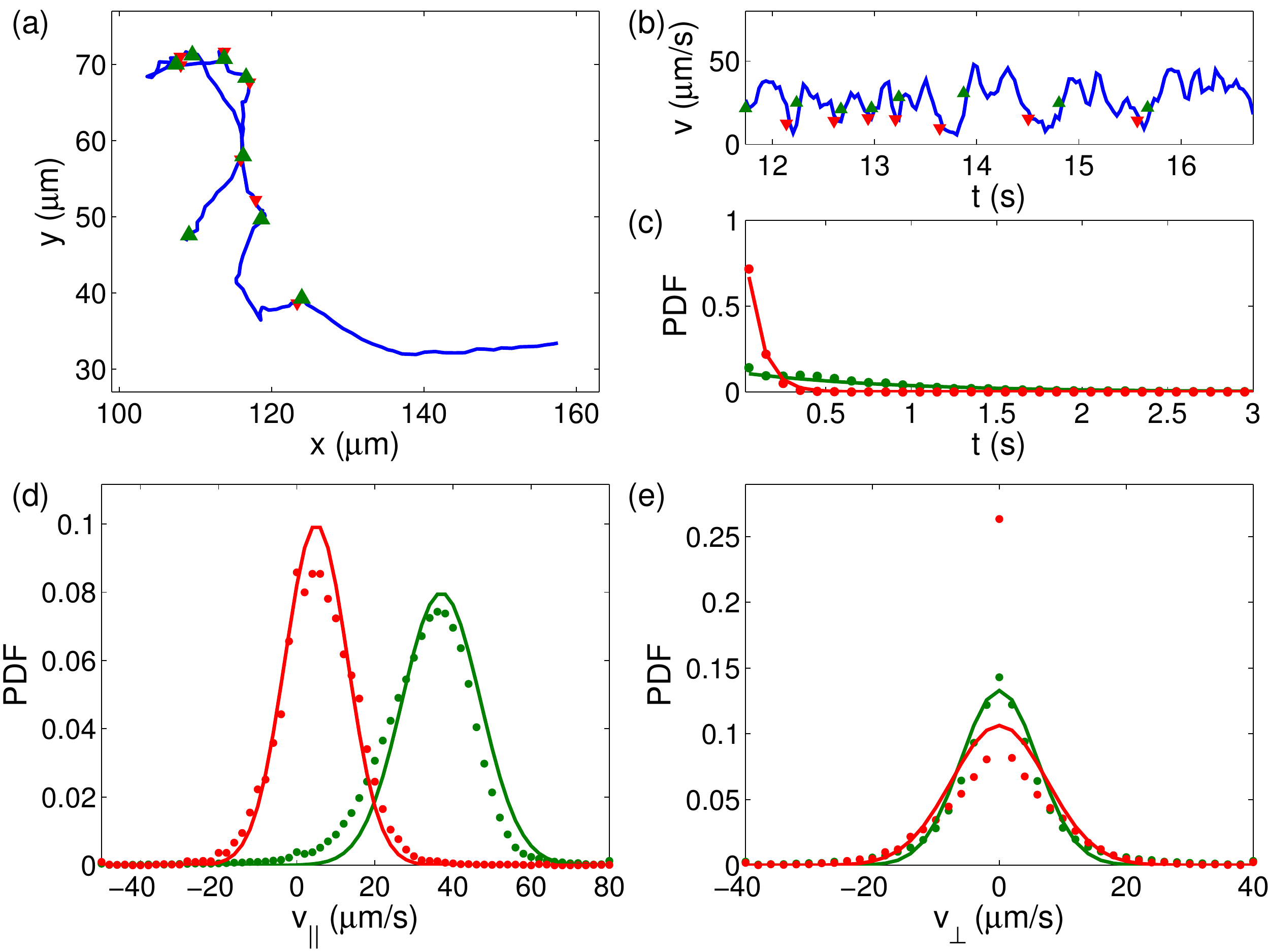}
\caption{(a) Typical experimentally obtained trajectory. Run and tumble starting points are respectively indicated with green and red marks. (b) Speed trace from the same trajectory in (a). (c) Experimental residence-time PDFs (dots) and best fitting exponential distributions (solid lines) for runs (green) and tumbles (red). (d) Experimental longitudinal velocity PDFs for runs (green dots) and tumbles (red dots), and the corresponding best fitting Gaussian distributions (green and red solid lines). (e) Same as in (d), but for transverse velocity.}
\label{Fig02}
\end{figure}

In order to have a point of reference, we started by analyzing the trajectories obtained from very low (about 0.01) obstacle-area-fraction videos. To do so, we computed the instantaneous speed and, as explained in the Materials and Methods section, we used it to classify bacterial swimming as either persistent (runs) or tumbling (tumbles). A typical trajectory is shown in Fig. \ref{Fig02}(a), with the run and tumble starting points indicated with green and red marks, respectively. The corresponding plot of speed vs. time is shown in Fig. \ref{Fig02}(b).

After measuring all the run and tumble resident times, we estimated the corresponding probability distributions (PDFs), and found that both of them can be fitted by exponential distributions. The experimental PDFs and the corresponding best fitting exponential distributions are plotted in Fig. \ref{Fig02}(c). The best fitting parameter values are tabulated in Table \ref{Tab01}. The observed one-order-of-magnitude difference between the run and tumble average duration times is consistent with previous reports \cite{Berg2004}. The difference between the average run time here observed and previously reported values could be due to solid surface interactions. See for instance the results in \cite{Lauga2006, Berke2008, Giacche2010, Drescher2011}, in which runs last longer due to hydrodynamic interaction with solid surfaces. 

To better understand the way bacteria move, the velocity at each trajectory step was decomposed into components perpendicular and transverse to the previous step (this was done separately for runs and tumbles), and we computed the corresponding PDFs for each one of these stages. I this case, we found that all the experimental PDFs are well fitted by Gaussian distributions. The experimental PDFs corresponding longitudinal velocity components for runs and tumbles are shown in Fig. \ref{Fig02}(d), together with the best fitting Gaussian distributions. The respective experimental and best fitting distributions for transverse velocity components are shown in Fig. \ref{Fig02}(d). The best fitting parameter values for all four PDFs are tabulated in Table~\ref{Tab01}.

We can appreciate in Figs.~\ref{Fig02}(d) and (e) that the longitudinal-velocity mean values corresponding to both runs and tumbles are positive and. Furthermore, as expected, longitudinal velocities are in general larger during runs than during tumbles. This is consistent with the fact that, on average, bacteria move along their longitudinal axis. Also observe that, according to both distributions, the probability of having negative longitudinal velocities is non negligible. These negative values make possible for bacteria to reverse or turn 180$^\circ$. This feature allows bacteria to scape bead traps by reversing their swim as described in \cite{Berg1995, Cisneros2006, Turner2000, Turner2010, Bianchi2015}. On the other hand, the transverse velocity distributions are symmetrical and they have zero mean values. However, the distribution corresponding to tumbles is wider than that corresponding to runs. As far as we understand, the symmetry of these distributions means that bacteria tumbles are unbiased, and the tumble transverse velocity distribution is wider because there is no net displacement in this motility mode.  

\begin{table}[htb]
\begin{ruledtabular}
\begin{tabular}{lcccc}
\multicolumn{5}{c}{Residence Time}\\
 & Run & Tumble  \\
$\tau$ & $1.2$ s & $0.1$ s \\
\hline
\hline
\multicolumn{5}{c}{Velocity Components}\\
&\multicolumn{2}{c}{$v_\parallel$ ($\mu m/s$) } & \multicolumn{2}{c}{$v_\perp$ $(\mu m/s$) }\\
& Run & Tumble &Run & Tumble\\
\colrule
$\mu$ & 37.84 & 5.99 & 0 & 0\\
$\sigma$ & 10.68 & 8.526 & 6.092 & 7.589\\
\end{tabular}
\end{ruledtabular}
\caption{Parameter values for the exponential distributions that best fit the run and tumble resitend-time PDFs, ans well as for the Gaussian distributions that best fit the longitudinal and transverse velocity PDFs corresponding to both runs and tumbles.}
\label{Tab01}
\end{table}

\subsection{Effect of increasing obstacle concentration}

As we can appreciate in Figs. \ref{Fig01}(a) and (b), bacterial trajectories display noticeable changes as the obstacle concentration increases. For instance, at low obstacle concentrations, bacterium swimming is more persistent than it is at higher obstacle concentrations. On the other hand, obstacle arrangement at high concentrations makes it possible to find several configurations such as: corridors, chambers, and inaccessible areas, all of which seem to affect bacterial motility. 

To investigate the effect that increasing obstacle concentration has upon bacterial motility, we recorded bacterial motility videos at different obstacle area fractions in the interval $[0.01, 0.4]$, recovered all the available bacterium trajectories, and analyzed them with the techniques introduced in the Materials and Methods section.

\begin{figure}[htb]
\includegraphics[width=.48\textwidth]{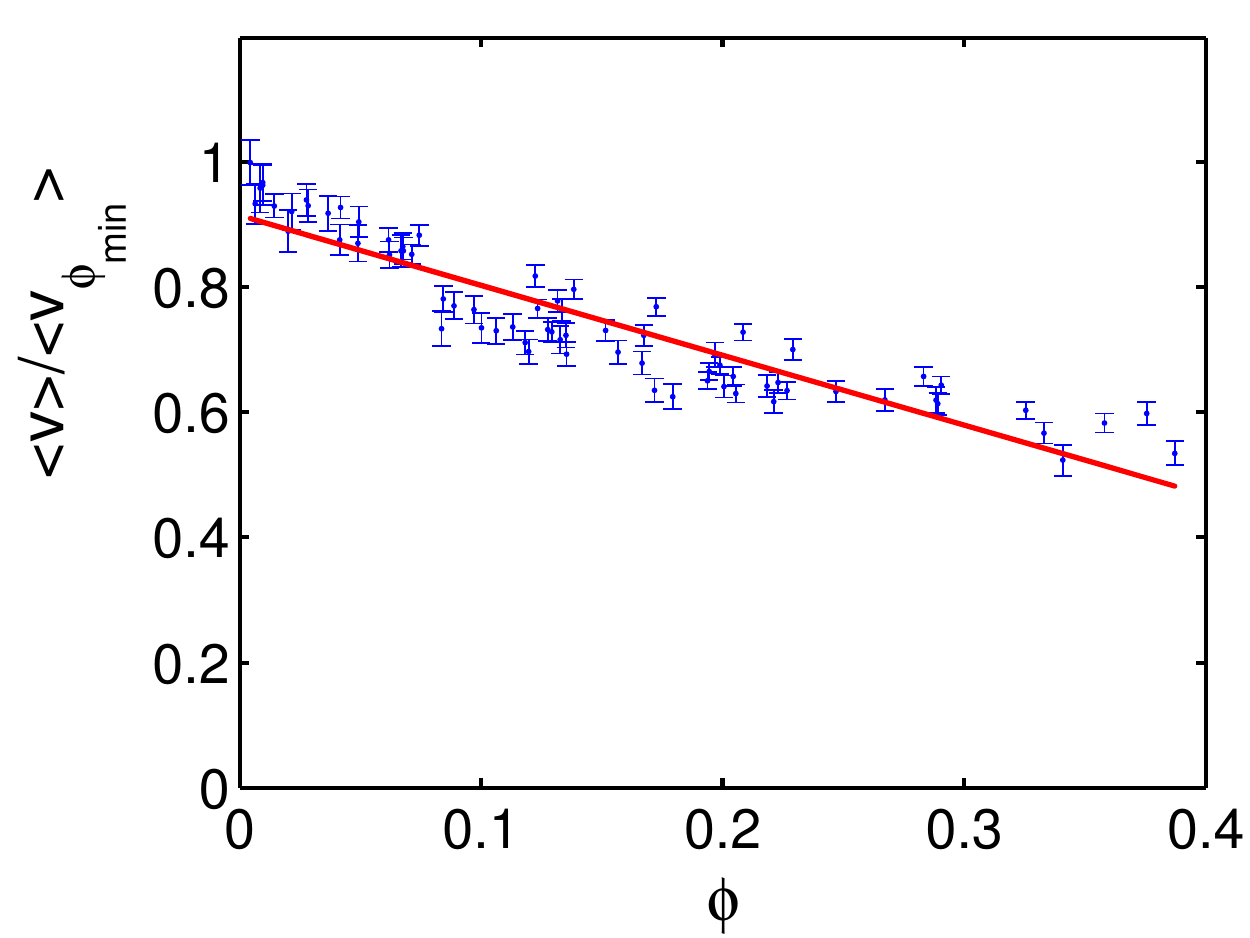}
\caption{Plot of the experimentally measured average bacterial speed (normalized to the value obtained at very low obstacle concentration) vs. the obstacle area fractions at every experiment.}
\label{Fig03}
\end{figure}

First of all, we computed the average bacterial speed for each experiment, and plotted the results vs. the obstacle area fraction ($\phi$) in Fig. \ref{Fig03}. Observe that such curve is well fitted by the following straight line:
\begin{equation}
  \langle v \rangle = \langle v \rangle_{\phi_{min}} (1 - m \phi),
  \label{Eq09}
\end{equation}
where $\langle v \rangle_{\phi_{min}}$ is the average bacterial speed at very low obstacle area fraction, and $m \approx 1.2215$. 

In what follows, we shall compare the results from experimental and simulated trajectories (simulated by means of the model described in the Materials and Methods section). To account for the previous paragraph results, we modified the model algorithm as follows. In step 6, if the bacterium is in the persistent motility mode (i.e. in a run), instead of randomly computing the longitudinal velocity component from a Gaussian distribution with the mean value reported in Table \ref{Tab01}, we took
\[
 \mu = \mu_0 (1 - m \phi),
\] 
with $\mu_0 \approx 37.84$ $\mu$m/s, and $m$ as defined above. For the standard deviation we employed the value reported in Table \ref{Tab01}. Finally, each simulation accounted for ten bacteria swimming in a $120 \times 160$ $\mu$m surface, with periodic boundary conditions. In Figs. \ref{Fig01}(c) and (d) we show a few representative simulated trajectories at low and high obstacle concentrations, respectively.

\begin{figure}[htb]
\includegraphics[width=.48\textwidth]{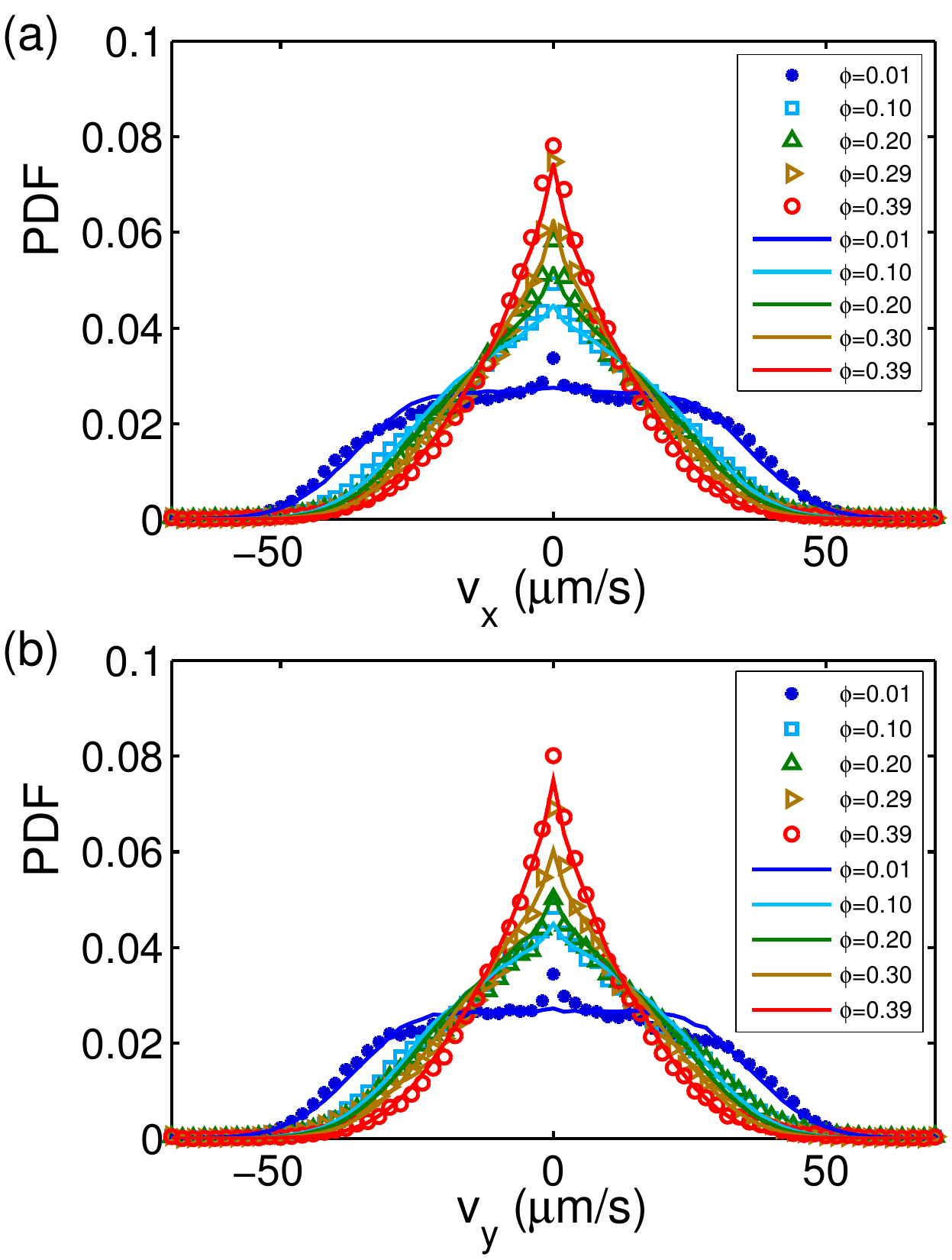}
\caption{Probability density functions for the bacterium horizontal ($v_x$) and vertical ($v_y$) velocity components, calculated from experimental (symbols) and simulated (solid lines) trajectories, at different obstacle area fractions ($\phi$):  blue, $\phi \approx 0.01$; cyan, $\phi \approx 0.1$;  green, $\phi \approx 0.2$, cyan, $\phi \approx 0.1$, light brown, $\phi \approx 0.29$, red, $\phi \approx 0.39$.}
\label{Fig04}
\end{figure}

In order to test for external fluxes and/or chemotaxis, we measured the horizontal ($v_x$) and and vertical ($v_y$) velocity components at every step from both experimental and simulated trajectories, and calculated the corresponding probability density function.  The results are shown in Fig. \ref{Fig04}. There, we can appreciate that all PDFs are symmetrical and that the $v_x$ and $v_y$ probability distributions are very much alike. To our consideration, these results allow us to disregard both external fluxes and chemotaxis in our experimental setup. Another interesting observation is the fact that the PDFs become wider as the obstacle area fraction ($\phi$) decreases. This last result is in agreement with the our previous appreciation that the average bacterium speed decreases as $\phi$ increases. Finally, we wish to emphasize that the PDFs computed from the simulated trajectories agree very well with the experimental ones. .

\begin{figure}[htb]
\includegraphics[width=.48\textwidth]{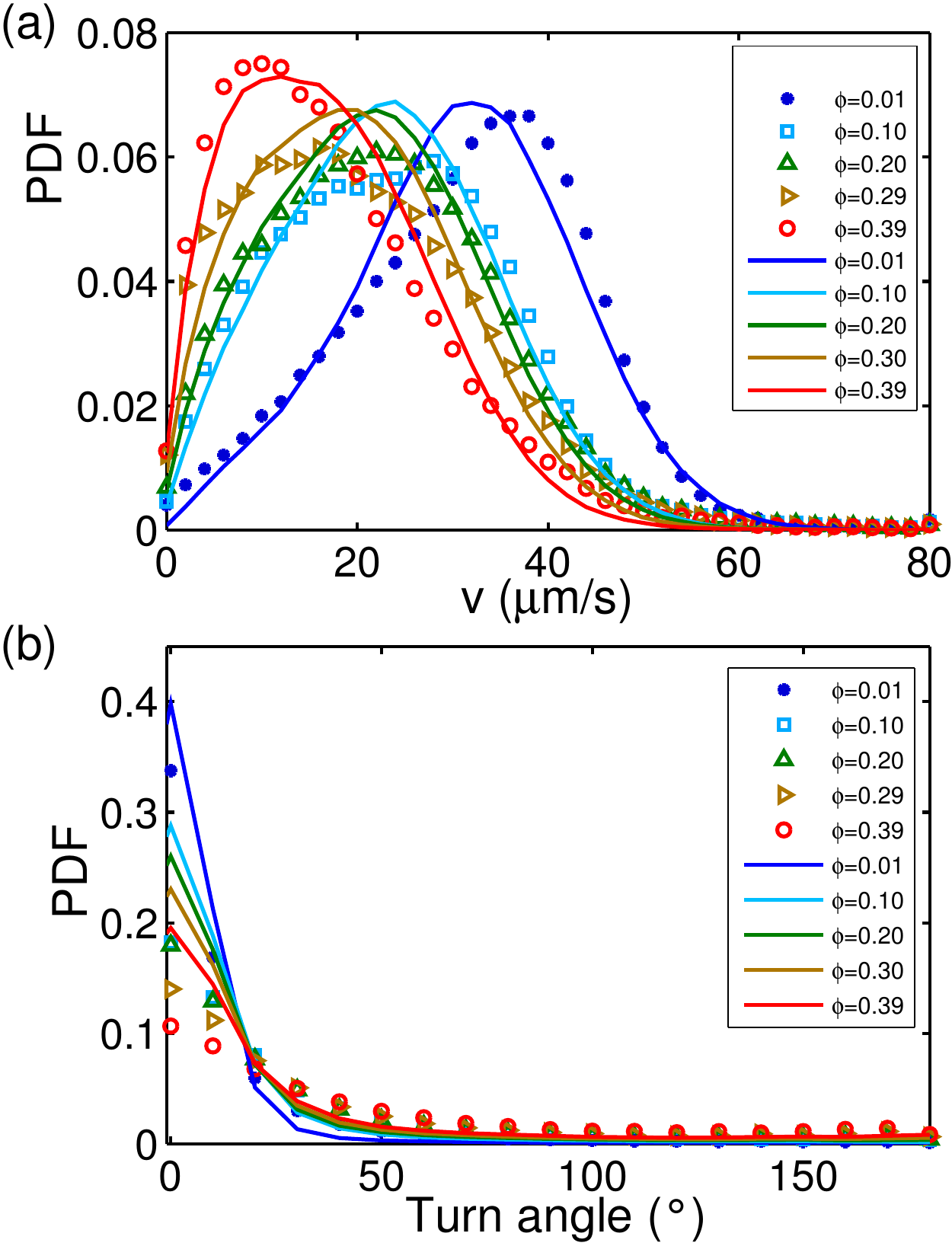}
\caption{(a) Speed probability density functions (PDFs) computed from experimental (symbols) and simulated (solid lines) trajectories at various obstacle area fractions ($\phi$):  blue, $\phi \approx 0.01$; cyan, $\phi \approx 0.1$;  green, $\phi \approx 0.2$, cyan, $\phi \approx 0.1$, light brown, $\phi \approx 0.29$, red, $\phi \approx 0.39$. (b) Turn-angle PDFs computed from experimental (symbols) and simulated (solid lines) trajectories at various obstacle area fractions. The color code is the same as in (a).}
\label{Fig05}
\end{figure}

After confirming that, in our experiments, bacterium swimming is unbiased, we decided to characterize how bacterial motility is affected by the the presence of obstacles at different concentrations. We started by taking the experimental and simulated trajectories, measuring the instantaneous speeds from all trajectories, and reckoning the corresponding probability density functions (one for each obstacle area fraction). The results are shown in Fig. \ref{Fig05}(a). Note that, once more, there is a good agreement between the simulated and the experimental results. Observe as well that all speed PDFs are unimodal, and that increasing obstacle concentration makes the position of PDF mode decrease. This is in agreement with our previous result that, on average, bacteria become slower as the obstacle concentration increases. On the other hand, we can observe that all PDFs are heavy tailed and in consequence, there is a non-negligible probability that bacteria move with speeds as large as 50 $\mu$m/s, even at the higher obstacle concentrations. We believe that this fact is associated to the existence (at high obstacle concentrations) of long corridors along which bacteria can swim in a mostly rectilinear fashion. 

The turn angle between consecutive steps is another helpful parameter to characterize complex trajectories, like the ones here obtained. Hence, we measured this parameter from all the experimental and simulated trajectories, computed the corresponding PDFs (one for every considered obstacle area fraction), and plotted the results in Fig. \ref{Fig05}(b). Observe that the turn-angle PDFs become wider and heavier tailed at the obstacle concentration increases. This means that, as expected, turn angles increase on average at higher obstacle concentrations.
On the other hand, when contrasting the experimental and the simulated results, we can observe that, invariably, the PDFs computed from the simulated trajectories render larges values at low turn angles, than the corresponding experimental PDFs. Moreover, this difference becomes more notorious at larger obstacle concentrations. Although definitely more work is necessary to explain these differences, we believe that thay can be due to bacterium-obstacle interactions not accounted for by our model.

\begin{figure}[htb]
\includegraphics[width=.48\textwidth]{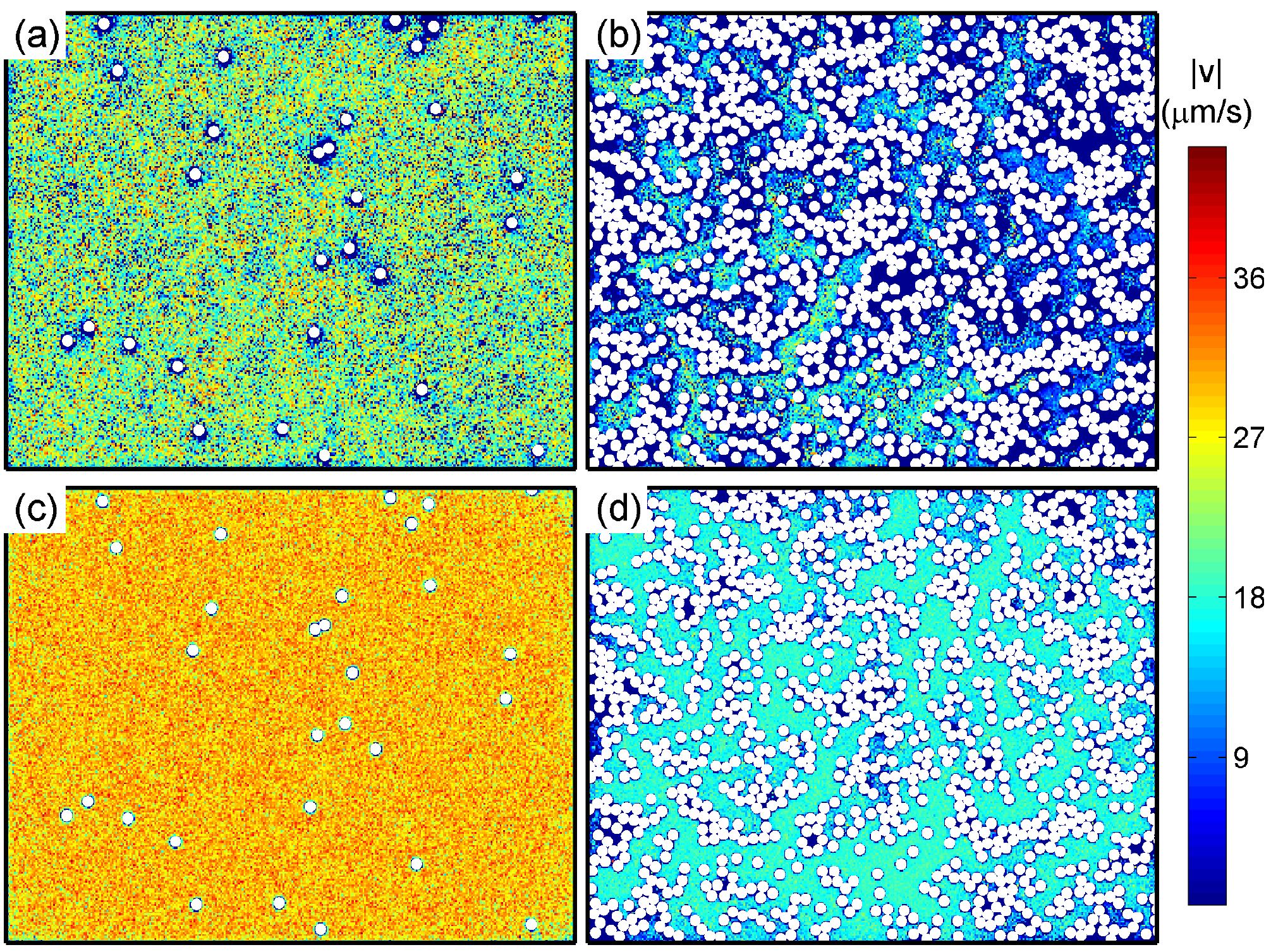}
\caption{Mean speed (averaged over time and over all trajectories in every squared micron) maps computed from experimental---(a) and (b)---and simulated---(c) and (d)---trajectories at different obstacle concentrations (Same experiments as those in Fig. \ref{Fig01}). The color code is given in the adjacent bar. Spherical beads are represented as white circles.}
\label{Fig06}
\end{figure}

To better understand the previous paragraph results, we measured the average bacterial speed (averaged over time and over all available trajectories) in every squared micron of the recorded field, and plotted the results in Fig. \ref{Fig06} for the experimental and the simulated trajectories. We can observe there that, in the experimental trajectories, the average bacterial speed decreases in the neighborhood of obstacles, and that this phenomenon is more notorious as obstacle concentration increase. For instance, when two or three obstacles are close together, the average bacterial speed  between them is even smaller than in the vicinity of isolated obstacles. Moreover, when larger obstacle concentrations lead to corridors, chambers and inaccessible areas (see Fig.~\ref{Fig06}(b)), bacterial speed in corridors and chambers is about 25 $\mu$m/s and 15 $\mu$m/s, respectively. That is, bacterial motility at high obstacle concentrations is severely affected but not completely stopped, and bacteria are still able to escape from traps and visit all the available space. In contrast, we can appreciate from Fig.~\ref{Fig06}(c) and (d) that, in the simulated trajectories, bacterial speed is pretty much homogeneous across the space not occupied by obstacles, even in the vicinity of single obstacles or between nearby obstacles. Furthermore, no difference in speed can be appreciated inside corridors and chambers at higher obstacle concentrations. In summary, these results corroborate that, as previously asserted, bacteria interact with obstacles in a complex way that is not accounted for in our model.

\begin{figure}
\includegraphics[width=.48\textwidth]{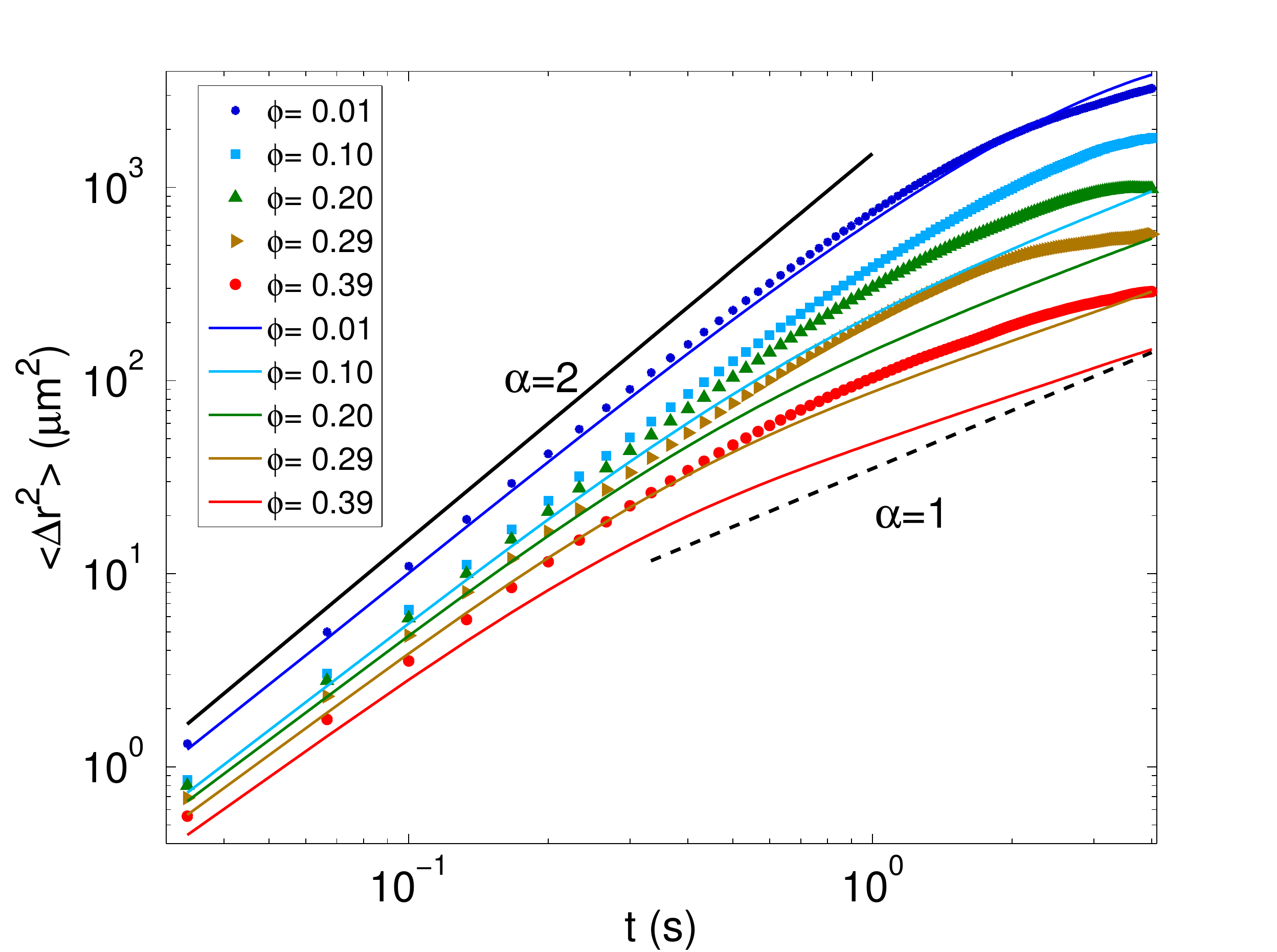}
\caption{Plots of mean squared displacement (MSD) vs. time, computed from experimental (symbols) and simulated (solid lines) trajectories, at different obstacle area fractions ($\phi$).}
\label{Fig07}
\end{figure}

Finally, we computed the mean squared displacement (MSD) from both the experimental and the simulated trajectories, and present the results in Fig. \ref{Fig07}. There, we can appreciate that, in both cases, the MSD slope is negatively correlated to the area fraction occupied by obstacles. Moreover, at short times, all MSD's denote super-diffusive motion, as expected for  auto-propulsive colloidal particles. Also expected is the fact that the duration of super-diffusive motion decreases as obstacle concentration increases; because the distance bacteria swim without interacting with obstacles decreases with obstacle concentration. Interestingly, the model results resemble the experimental ones during the super-diffusive period for all obstacle concentrations. But only at very low obstacle concentrations there is a good agreement between the simulated and the experimental MSDs for longer times. Once again, we believe that bacterium-obstacle interactions missing in our model are responsible for these differences. 

\section{Conclusions}

Microorganisms constitute a large proportion of total Earth's biomass, and a many of them live in porous media \cite{Whitman1998}. Henceforth, understanding how bacterial motility is affected by porous conditions is important for research fields like: soil microbiology, bioremediation, microbial pathogenesis and others. Bacterial motility in micron and sub-micron constricted spaces has been studied with ideal geometries \cite{Mannik2009, Berdakin2013, Ping2015, Raatz2015, Jabbarzadeh2014}. In this work we have generated a more complex and realistic environment that simulates a quasi-two-dimensional porous media \cite{Davit2013}, and studied how \textit{E. coli} motility is affected by decreasing porosity conditions.

To characterize bacterial motility we recorded motile bacteria, recovered the trajectories by means of image analysis techniques, and statistically analyzed trajectory parameters like: instantaneous velocity, turn angle, mean squared displacement and speed map.  We carried out experiments in which the area fraction occupied by obstacles spans from 1 to 39 $\%$.

From the obtained results, we hypothesized that bacterial movement can be described  as follows:
\begin{itemize}
  \item \textit{E. coli} presents two different motility modes: runs and tumbles, and a given bacterium randomly and spontaneously switches between these two modes.
  
  \item Bacterial swimming can be viewed as phenomenological random walk in which the longitudinal and transverse components of each step (as referred to the previous step), obey characteristic probability density functions for every motility mode. 
  
  \item Bacterial average speed linearly decreases as the obstacle concentration increases.
  
  \item Bacteria only interact with obstacles through collisions.
\end{itemize}
 All these characteristics were taken into account to build the model described in detail in the Materials and Methods section. This model was employed to simulate bacterial trajectories at different obstacle concentrations, and to compare them with the experimentally obtained trajectories.

Given that the speed PDFs obtained from the simulated trajectories match those of the experimental trajectories for all obstacle concentrations, while the results from simulated trajectories for the turn angle PDFs and the mean squared displacement qualitatively agree with the experimental ones, we conclude that our model can be regarded as a sound semiquantitavie description of bacterial motility in a quasi-two-dimensional porous medium. Concerning the quantitative differences observed between the experimental and the simulated  turn-angle PDFs and mean squared displacement curves, or analyses suggest that they may be due to bacterium-obstacle interactions not accounted for by our model. For instance, these extra interactions may be of hydro-dynamic nature, or due to physical contact between the bacterium flagella and the spherical beads \cite{Turner2000, Turner2010, Davit2013, Lushi2014}. Answering this questions demands future experimental and mathematical-modeling work



\bibliography{BibEcoli}

\end{document}